\documentclass[aip,rsi,reprint]{revtex4-1} 

\usepackage{amsmath}
\usepackage{graphicx}

\DeclareMathOperator{\e}{{\displaystyle e}}
\DeclareMathOperator{\ber}{{\displaystyle ber}}
\DeclareMathOperator{\bei}{{\displaystyle bei}}
\newcommand{\rd}{\sqrt{2}}

\graphicspath{{Figs/}}

\begin{document}
\title{A room-temperature alternating current susceptometer - Data analysis, calibration, and test}
\author{M. Alderighi}
\author{G. Bevilacqua}
\author{V. Biancalana}
\author{A. Khanbekyan}
\affiliation {CNISM, CSC and DIISM, Universit\`a di Siena,  via Roma 56, 53100 Siena, Italy.}

\author{Y. Dancheva}
\author{L. Moi}
\affiliation {CNISM, CSC and DSFTA, Universit\`a di Siena,  via Roma 56, 53100 Siena, Italy.}

\begin{abstract}
An AC susceptometer operating in the range of 10~Hz to 100~kHz and at room temperature is designed, built, calibrated and used to characterize the magnetic behaviour of coated magnetic nanoparticles. Other weakly magnetic materials (in amounts of some millilitres) can be analyzed as well. The setup makes use of a DAQ-based acquisition system in order to determine the amplitude and the phase of the sample magnetization as a function of the frequency of the driving magnetic field, which is powered by a digital waveform generator. A specific acquisition strategy makes the response directly proportional to the sample susceptibility, taking advantage of the differential nature of the coil assembly. A calibration method based on conductive samples is developed.

\end{abstract}

\date{\today}



\maketitle


\section{Introduction}

AC susceptometry constitutes a powerful tool for research into the characterization of magnetic materials. The construction of  simple, versatile, automatized and user-friendly apparatuses is of interest for researchers in many areas in which the magnetic properties of weakly magnetized materials need to be investigated.  Some apparatuses use highly homogeneous magnetizing fields, taking advantage of the Helmholtz configuration \cite{chen_04}, but most of them (including commercial ones) use simple solenoids. Quite often the real and the imaginary components of the susceptibility (accounting for magnetic energy storage and for power dissipation, respectively) are extracted with the help of a lock-in amplifier. Modern lock-in amplifiers are excellent devices for noise rejection, acquisition and the analysis of signals. On the other hand, the spurious  signals that typically emerge in susceptometric measurements, due to minor imperfections, may affect the lock-in output unless sophisticated circuits are inserted to nullify them before the phase-sensitive detection. Such circuits can be expensive, require delicate adjustments and may introduce other systematic errors. 

We present here a solution based on the direct digitization of raw data, which are subsequently numerically elaborated in an off-line analysis so as to cancel spurious signals, using relatively inexpensive instrumentation. The complex susceptibility as a function of the frequency is derived using a simple calibration method.

Among the wide range of materials requiring magnetic characterization, we concentrate on superparamagnetic nanoparticles embedded (or to be embedded) in an organic matrix 
(e.g. hydrogels). It has been suggested that magnetic nanoparticles (MNPs) surface-functionalized with biological probes can be attached to cancer cells for selective killing and used for innovative therapies such as controlled drug delivery \cite{hergt_06, mamiya_11, arruebo_07}, or in biology as magnetic labels \cite{grossman_04}.

MNPs subjected to an alternating magnetic field can exchange energy with it. With a suitably selected  AC magnetic field amplitude and frequency,  the particles can generate enough heat to damage or kill the surrounding cells, while avoiding collateral damage to neighbouring tissues. A serious problem with these magnetic hyperthermia applications is that the field applied induces eddy currents in the exposed tissues and organs. This may cause a significant transfer of energy with consequent collateral effects, thus thwarting the selectivity associated with the MNP methods. 

In all the above-mentioned applications and wherever AC magnetic field manipulation of MNPs is required, it is important to know the size of the MNPs, their statistical distribution and the near-dc susceptibility value, which make it possible to estimate the rate of the energy transferred to the sample, or rather the specific absorption rate at a given AC magnetic field frequency. The response of the MNPs is, in fact, strongly dependent on the dominant relaxation mechanism (Ne\'el and Brown relaxations may occur with superparamagnetic materials, the second prevailing for larger MNP sizes) \cite{hergt_98}, whose typical time response dramatically depends  on the size and - in the case of the Brown relaxation - on the hydrodynamic radius of the particle. This marked dependence makes preliminary experimental characterization of the resulting susceptibility advisable in all applications where a significant field-MNP energy transfer is required.

In many cases detailed magnetic response studies are performed to analyse the temperature dependence of the magnetic susceptibility, with particular emphasis on its behaviour at cryogenic temperatures \cite{couach_85}. In other cases (including the biomedical field of interest) the temperature is not an adjustable parameter, and the optimal frequency and amplitude are the only quantities to be determined.
Commercial susceptometers are available (often coupled with cryogenic systems and thus rather expensive) and home-made setups are reported in the literature \cite{laurent_08,dumelow_01}. Home-made setups offer the advantage of adapting the sensor volume to the sample size. The geometrical constraints, together with the coil design, also have implications for the accessible frequency range. In conclusion, the use of {\it ad hoc} apparatuses, such as home-made ones specially designed for a given sample size and frequency range, helps in optimizing susceptibility measurements. 

Our setup works in a relatively weak driving field (in the $\mathrm{1~mT}$ range and lower) and is devoted to the study of magnetization phenomena occurring in a linear regime (far from saturation). In this paper we present a simple design for a room-temperature AC susceptometer using basic modules commonly available in scientific laboratories. The apparatus could be coupled with an additional coil to provide a stronger dc or low-frequency range field, in order to investigate the nonlinear terms of the susceptibility. A detailed description of the susceptometer calibration procedure is provided, which permits simultaneous accurate measurements of both the real and the imaginary components of the  susceptibility.  The calibration procedure is based on the detection of eddy currents induced in large aspect-ratio Cu rod samples \cite{chen_99}. These kinds of samples have a theoretically known diamagnetic behaviour, so that the experimentally measured response can be compared with a theoretical model. 

The theoretical expression includes some special functions that do not allow for immediate analytical evaluation. We report (in an Appendix) the mathematical formulas necessary to calculate the theoretical response of cylindrical conductive samples, whose diameter can  be selected in order to achieve precise calibration in the desired frequency range. Lastly, a few susceptometric measurements are presented and discussed, showing the performance of the susceptometer when used to characterize samples containing MNPs of different sizes.

\section{Experimental setup} \label{sec:setup}

\begin{figure}[htbp] \centering
\vspace{12pt}
\includegraphics [width=8cm] {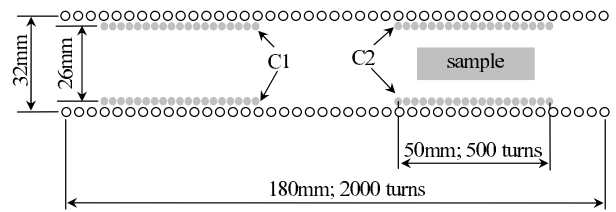}
\caption{Schematic of  the coil assembly (in scale).}
\label{setup}
\end{figure}

The susceptometer is based on a simple geometry coil assembly. The assembly is composed of three coaxial, cylindrical coils: a large one used to apply a homogeneous field, and two (identical) smaller ones used to pickup an electric signal via the Faraday induction effect
(coils C1 and C2 in Fig.\ref{setup}).
All the coils are made of a single layer copper wire wound around thin plastic cylinders. Their loops are  wound accurately so as to make them very regularly distributed. The wire diameter of the main and the pick-up coils is chosen in accordance with the frequency range to be investigated. Parasitic capacitance causes unwanted resonances, preventing the use of thin wires (too closely spaced loops) when higher frequencies need to be investigated. 

The pickup coils fit inside the cylinder of the main coil. The volume of the pickup coils should be such as to provide maximum coupling with the sample volume, in order to achieve maximum sensitivity. The setup presented in Fig.\ref{setup} is chosen in order to accommodate MNPs either dry or dispersed in about 4~ml of liquid. 
The main coil is supplied by a high quality waveform generator, which  delivers a 10~$V_{pp}$  sinusoidal signal on a 50~$\Omega$ load: it delivers 20~$V_{pp}$ 
to the coil at higher frequencies, where its reactance largely exceeds the output impedance of the generator. The magnetic field inside the main coils is of the order of 
1~mT at the lowest frequency (10~Hz) and drops down to $20~\mathrm{\mu T}$  at 100~kHz. In all cases it is far below the saturation limit of the samples considered in this paper.

The signals produced by the pickup coils are digitized by means of a DAQ card (16 bit resolution, 1.25~MS/s acquisition rate).  The two pickup coils are placed inside the main coil, taking care to position them in regions where a sufficiently homogeneous driving field is generated, i.e. at a distance of about one diameter from the ends of the main coil. 

We now describe the measurement method and, more specifically, the approach used to calibrate the susceptibility and to counteract systematic errors. The naive idea for measuring the susceptibility $\chi$ of a material is to insert a sample in a homogeneous field and measure the field variation caused by the sample magnetization. The most direct method for measuring a time-dependent (harmonically oscillating) field is to use a pickup coil. The insertion of a sample in a pickup coil placed in a homogeneous oscillating field would cause a relative variation in the induced electromotive force (EMF):
\begin{equation}
\frac{\Delta\epsilon}{\epsilon}=\frac{\epsilon^0+\epsilon^{sample}}{\epsilon^0}=\chi\frac{V_S}{V_C} ,
\label{eq:deltaepsilonrel}
\end {equation}
where $\epsilon^0$ and  $\epsilon^{sample}$ are the induced EMF in the absence and in the presence of a sample, respectively; $V_S$ is the volume of the sample and $V_C$ is the volume of the pickup coil. In fact, the magnetic field changes from $B_0$ to $B_0(1+\chi)$ in the volume occupied by the sample, resulting in a modified time-dependent magnetic flux across the pickup coil. The estimate \eqref{eq:deltaepsilonrel} is correct when an infinite volume sample in an infinite pick-up coil is considered. Actually the ratio $V_S/V_C$ is a rough estimate of a coupling factor $\alpha$, which describes the coupling and the filling factors and (what is important) is set only by geometric values
\begin{equation}
\frac{\Delta\epsilon}{\epsilon}=\alpha\chi .
\label{eq:deltaepsilonrelalpha}
\end {equation}
Small values of $\chi$ and possibly small values of $\alpha$ make the  relative variation of the EMF measured difficult to detect and digitize correctly. A straightforward solution to improve the digitization resolution is to subtract the background EMF (that of the empty pickup coil) by performing differential measurements. This can be done by using two identical pickup coils (see Fig.\ref{setup}) and measuring the differential signal when inserting the sample in the pickup coil C1 and alternately in C2, thus obtaining:
\begin{eqnarray}
\Delta\epsilon_1=\left[\epsilon_1^0(1+\alpha_1\chi)-\epsilon_2^0\right]  \mbox{and} \nonumber \\
\Delta\epsilon_2=\left[\epsilon_2^0(1+\alpha_2\chi)-\epsilon_1^0\right],
\label{diff:signal}
\end{eqnarray}
respectively. Adding the two quantities measured, a signal  free of background EMF (differential signal) $(\epsilon_1^0 \alpha_1+\epsilon_2 ^0\alpha_2) \chi$  is promptly made available. Extraction of the $\chi$ value now requires precise evaluation of the parenthesis. The relative position of the two pickup coils is finely adjusted in such a way as to minimize the amplitude of the differential signal measured when no sample is inserted and the main coil is driven at an intermediate frequency and at a strong magnetic field. Achieving a good balance (minimal differential signal) makes it possible to achieve  a good digitization of the signals produced by the presence of the sample. With the availability of two nearly identical pickup coils, and positioning the sample inside them accurately, the assumption that the coupling factors are equal $\alpha_1=\alpha_2=\alpha$ can be made, so that the necessary normalization factor becomes $\alpha(\epsilon_1^0+\epsilon_2^0)$. Further simplification based on the hypothesis that  $\epsilon_1^0=\epsilon_2^0$ would not be appropriate, because small imperfections (such as parasitic capacitance and consequent spurious resonances) make such EMFs different, mainly in terms of their relative phase and at  higher frequencies, where the effects of uncontrolled resonances become increasingly important. For the same reason,  the position of the pickup coils that gives minimum differential signal in the balancing procedure depends on the frequency applied.
Nevertheless, also thanks to the signal extraction method (see below), we verified that such imperfections do not prevent accurate measurements. We also verified that the susceptibility measured does not depend on the frequency at which the balancing procedure is performed.

We followed a four-step measurement procedure to determine susceptibility, based on two differential measurements (with the sample in C1 and C2, respectively) and two absolute measurements of the EMF induced in the empty coil C1 and C2, respectively. Thus the quantities measured are: $\Delta\epsilon_1, \Delta\epsilon_2,  \epsilon_1^0$, and $\epsilon_2^0$. The $\alpha\chi$ value is then estimated as  $(\Delta\epsilon_1+\Delta\epsilon_2)/(\epsilon_1^0+\epsilon_2^0)$. The remaining task is to determine $\alpha$. In addition, as all the quantities considered are actually functions of the angular frequency $\omega$, fine calibration of the response at different frequencies is necessary.

Most of the setups reported in the literature are based on a lock-in amplifier (see, for example \cite{chen_05, goldfarb_84}) which is used both to provide the driving field (in some cases with an amplification stage to improve the field strength) and to reveal the phase and amplitude of the induced EMF. In our case, we use the raw data directly and perform an {\it ad hoc} analysis. This choice  helps to reduce the complexity and the cost of the whole setup, and facilitates the task of optimizing the data acquisition and analysis.

\begin{figure}[htbp] \centering
\includegraphics[width=8cm]{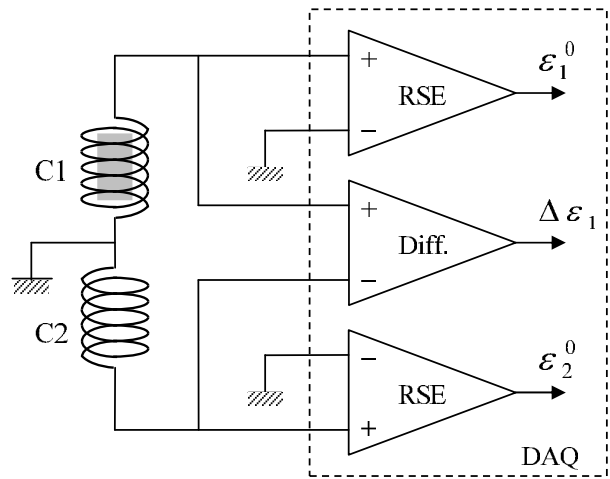}
      \caption{Schematic of the DAQ connections. A sequence of four equal frequency scans is performed and for each scan a specific single signal is acquired at each frequency. The sample is placed alternately in the coil C1 or C2. When the sample is in C1, as sketched in the figure, the central (differential) input provides $\Delta\epsilon_1$ and the lower (RSE) converter provides $\epsilon_2^0$. When it is in C2, the upper (RSE) converter provides $\epsilon_1^0$ and the central one provides $\Delta\epsilon_2$.}
				\label{sushidaq} 
\end{figure}

We use a commercial ADC converter (National Instruments, NI-PCI-6250). This  provides multiple input conversion, however we prefer to use repeated single input conversions to avoid any cross-talk effects and  phase errors  introduced by the multiplexed digitization. The two pickup coils are connected in series and grounded at their connection (see Fig.\ref{sushidaq}). The signal produced by each single coil can thus be digitized in a referenced single-ended (RSE) configuration, while the small series signal is digitized in a pure differential configuration, which permits optimal rejection of the common mode noise. In all cases the digitization is triggered by the TTL reference signal produced by the waveform generator.

The voltage range of the 16bit-ADC card can be set from +/-100mV up to +/-10V, leading to a minimum digitization uncertainty of   $\delta \epsilon = 15 \mu$V, which is the dominant error source. In accordance with Eq.\ref{diff:signal}, this corresponds to an instrumental uncertainty of $\delta\chi = \delta \epsilon/ (\alpha \epsilon_0) \approx 1.2 \times 10^{-4}$,  calculated for a 1~ml sample (filling factor 3.7\%),  and for  $\omega  B = 12$ T\,rad/s, as in the case of our driving field. The averaging procedure may improve this value, while the presence of parasitic capacitance at high frequencies and coil resistance at low frequencies can worsen it.

During the frequency scan  a given number of periods of the signal (N) are recorded for each step. The acquisition rate is kept at the maximum, as long as the data buffer can contain the whole data set. Otherwise the rate is reduced so that the N periods fit in the data buffer. Multiple traces are acquired for each frequency step in order to further decrease the noise by means of trace averaging.

The choice of N (usually $\geq 5$) is made on the basis of empirical evaluation of the performance of the procedure used for amplitude and phase extraction. This is done using a numerical evaluation based on the analysis of windowed discrete Fourier transform (DFT), following the procedure described in Refs. \cite{grandke_83,lanczos_56}.

A sample susceptometric analysis starts with the selection of the frequency scan (a logarithmic discrete sweep, for which the operator sets the interval and the number of points) and the voltage applied to the main coil. The  measurement consists of four steps:
\begin{itemize}
\item
The sample is placed in the C2 coil and the EMF in C1 is measured  to evaluate ($\epsilon_1^0$) ($\epsilon_1^0$ is the complex number of which the above-mentioned procedure calculates modulus and phase).
\item
The EMF in the series ($\Delta\epsilon_2$) is evaluated.
\item
The sample is placed in the C1 coil and the EMF in C2 ($\epsilon_2^0$) is evaluated.
\item
The EMF in the series ($\Delta\epsilon_1$) is evaluated.
\end{itemize}
At the end of the four frequency scans, the four complex arrays are combined (see Eq.\eqref{diff:signal}) and rescaled by a calibration factor $g(\omega)$, which is evaluated from the linear interpolation of a calibration file previously saved to disk and produced according to the analysis described in Sec.\ref{calibration}

\section{Calibration}
\label{calibration}
A simple but very effective procedure to achieve accurate calibration (of both the coupling factor $\alpha$ and the frequency response) is based on the use of theoretically known susceptibility samples. As known \cite{chen_99, chen_05, chen_10}, cylindrical conductive samples with large aspect ratios (i.e. having a length greatly exceeding the diameter) serve this purpose well. We employ copper samples with diameters selected in order to match the range of interest of the susceptometer's frequency domain.

Sets of Cu wires, with diameters in the $1\div5$~mm range and a length of 40 mm, are used for calibration. For each diameter a variable number of Cu wires are used in order to get  roughly constant volumes, comparable to the volume of the sample of unknown susceptibility to be analyzed. The aim is to: \textit{(i)} determine the filling factor $\alpha$ (matching the resulting curves at the lower frequencies); \textit{(ii)} correct the frequency response of the pick-up coils (occurring at higher frequencies); and \textit{(iii)} validate the whole measurement procedure.
The $\chi$ value for a Cu cylinder of radius $r$ and length largely exceeding $r$ (in order to make the infinite length approximation feasible) can be calculated using:
\begin{equation}
\chi=\frac{2J_1(z)}{zJ_0(z)}-1
\label{eq:chi_rame}
\end {equation}
where $J_k $ is a Bessel function of the $k^\mathrm{th}$ order,
\begin{equation}
z=r\sqrt{-i\sigma\mu_0\omega}=\sqrt{-i\Omega}
\end {equation}
and $\omega=2\pi f$, $\sigma$ and $\Omega$ are the angular frequency, the material conductibility, and the dimensionless frequency defined as $\omega/(r^2\sigma\mu_0)^{-1}$, respectively \cite{landau_84, chen_99}. Eq.\eqref{eq:chi_rame} is the result of a classical eddy-current model.

\begin{figure}[htbp] \centering
\vspace{12pt}
\includegraphics [width=8cm] {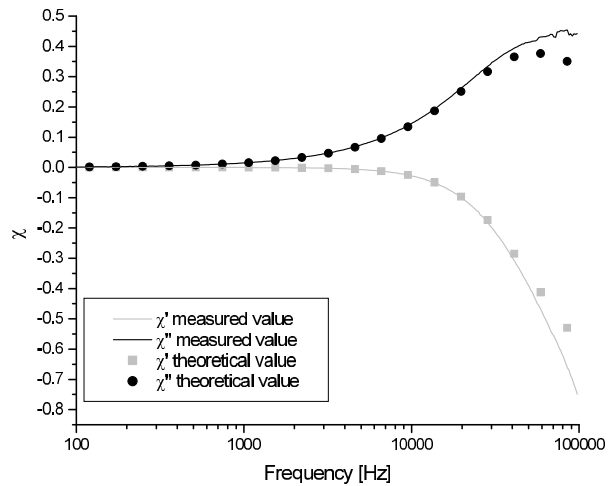}
\caption{Theoretical and experimental data used for calibration of the susceptometer. As a sample 20 wires of Cu, 1~mm in diameter and 40~mm in length, are used for
$\alpha g(\omega)$ determination.}
\label{rame1mm}
\end{figure}

The first task \textit{(i)} is accomplished and a reasonably good agreement is observed between the theoretical  [$\chi_{th}=\chi^{\prime}_{th}(\omega)-i\chi^{\prime\prime}_{th}(\omega)$] (see the Appendix) and experimental [$\chi_{exp}=\chi^{\prime}_{exp}(\omega)-i\chi^{\prime\prime}_{exp}(\omega)$] curves, using $V_S/V_C$ as an estimate of the coupling factor $\alpha$. Fig.\ref{rame1mm} shows a good rough agreement between the theoretical and experimental values, corrected with the filling factor only. It can be seen that  a discrepancy takes place at nearly 40~kHz due to the fact that the pick-up coils are not perfectly identical. The second task \textit{(ii)} can be accomplished by producing a complex calibration curve, reporting the ratio $g(\omega)=\chi_{th}/\chi_{exp}$ as a function of $\omega$, used to rescale the measurements afterwards.

\begin{figure}[htbp] \centering
\vspace{12pt}
\includegraphics [width=8cm] {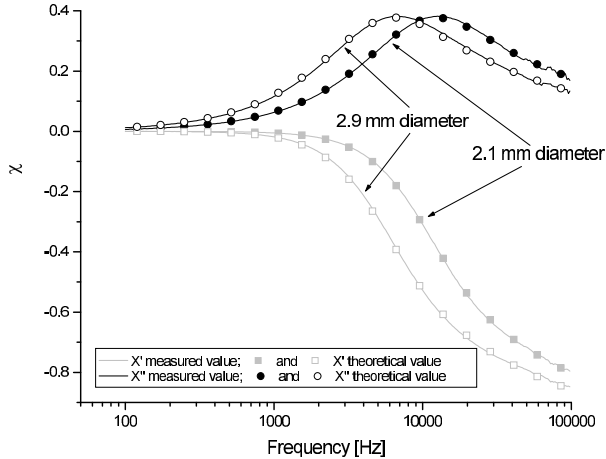}
\caption{$\chi ^\prime $ and $\chi ^{\prime \prime} $ theoretical and experimental values for 5 wires  of 2.1~mm in diameter and for 3 wires  of 2.9~mm in diameter. Both Cu samples are 40~mm in length.}
\label{rame23mm}
\end{figure}

The quality of the calibration of the susceptometer was tested using other Cu wire samples of different diameters. Fig.\ref{rame23mm} shows excellent agreement between the experimental value and the value calculated using Eq.\eqref{eq:chi_rame}, when taking into account the small difference in the sample volumes. Appreciable discrepancies, due to the inappropriate assumption of an infinite length Cu cylinder, take place at diameters above 5~mm.

\begin{figure}[htbp] \centering
\vspace{12pt}
\includegraphics [width=8cm] {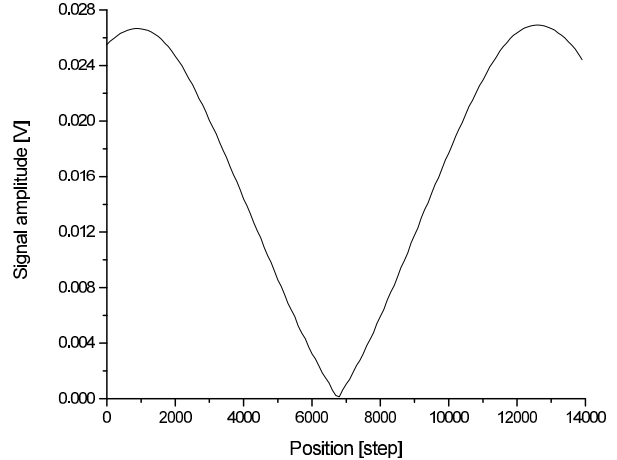}
\caption{The differential signal amplitude ($\Delta\epsilon_1+\Delta\epsilon_2$) as a function of the motor steps used to determine the correct positioning of the sample in the two pick-up coils.}
\label{positioning}
\end{figure}

The Cu calibration wires are used for fine sample positioning as well. The whole measuring procedure, described above, gives reliable results when $\alpha_1=\alpha_2=\alpha$ is  fulfilled and the signal-to-noise is optimized, provided that $\alpha$ is maximized,  i.e. provided that the sample is accurately positioned inside each pickup coil. To this end the setup includes a numerically controlled motorized positioner, for which a calibration procedure was developed. Prior to starting a measurement, it is possible to determine the optimal positions that maximize the coupling factors $\alpha_1 $ and $\alpha_2 $ and to verify that those factors are effectively equalized. For this purpose, the main field is set to oscillate at a fixed intermediate frequency, then position $x$ of the copper sample is scanned along the whole length of the main coil. The differential signal recorded $\Delta\epsilon_1+\Delta\epsilon_2$ is registered as a function of the sample displacement (see Fig.\ref{positioning}). 
The correct positions of the sample inside the C1 and C2 coils are those that produce the maximum amplitude. These maxima are determined using a single tone extraction procedure  \cite{grandke_83, lanczos_56}  and correspond to the maximum coupling values $\alpha_1 $ and $\alpha_2$, respectively.

\section{Applications}

Susceptibility measurements contain valuable information about the ac magnetic field manipulation capacity of a given material. Important information, such as the monodispersibility of the nanoparticles, their hydrodynamic diameter, and the relative amount of superparamagnetic nanoparticles free to move can be extracted.

\begin{figure*}[htbp] \centering
\vspace{12pt}
\includegraphics [width=12cm] {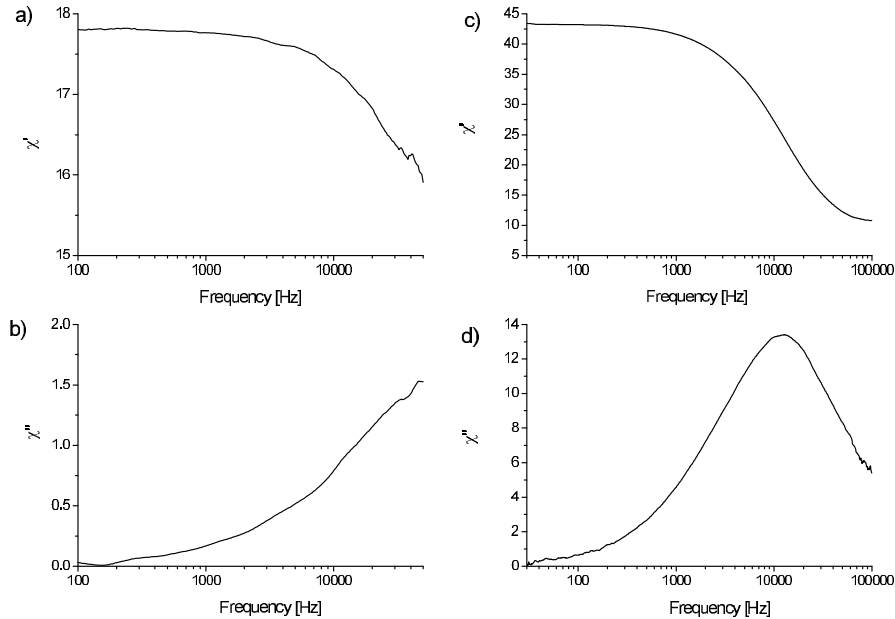}
\caption{$\chi^\prime $ and $\chi^{\prime\prime} $ value for a sample (2~ml; 13.8~mg of $\mathrm{Fe_3O_4}$) containing MNPs of 20~nm in diameter (plots a) and b)) and  25~nm in diameter (plots c) and d)).}
\label{mnp}
\end{figure*}

We measured the susceptibility of samples containing superparamagnetic particles dissolved in deionized water. The samples, provided by Ocean NanoTech, contain 5~mg/ml of iron in the ferrofluid. If it is  assumed that the particles are mainly made of $\mathrm{Fe_3O_4}$, the concentration of magnetite in the ferrofluid is about 6.9~mg/ml. The core diameter of the mono-dispersed nanoparticles is 20~nm, with a hydrodynamic size of 23.45~nm (standard deviation of 5.56~nm). The susceptibility measured is presented in Fig.\ref{mnp}. The maximum $\chi ^\prime$ value, as well as the frequency dependence of both $\chi^\prime$ and $\chi^{\prime\prime}$, differ from the theoretical value, but are in perfect agreement with the results presented in \cite{andersson_13}. 
As it is known (see for example \cite{hergt_98, kotitz_95}), at low frequencies the MNPs show  superparamagnetic behaviour, and at frequencies such that $\omega\tau>$1 (where 
$\tau$ is the measuring time) relaxation loss of energy takes place.  With a core diameter of 20~nm (anisotropy constant K=$1.35\times10^4~\mathrm{ J/m^3}$) and a hydrodynamic diameter of 30~nm, Brownian relaxation is dominant and determines a relaxation frequency of about 15~kHz, which is much lower than the experimentally measured frequency (see Fig.\ref{mnp}). 

However, the MNPs with a 25~nm core diameter (from the same provider) show a $\chi^{\prime\prime}$ maximum in very close correspondence to the theoretical value of  nearly 12~kHz (again determined by the Brownian relaxation), considering  a hydrodynamic diameter of 33~nm (the data sheet specifications provided by the producer declare a hydrodynamic diameter of 31.90~nm with a standard deviation of 7.04~nm).

A large discrepancy is obtained between the theoretically evaluated (dc) susceptibility and the measured one at low frequencies, with both samples. With a 20~nm core diameter $\chi^\prime$ can be calculated to be about 270 (saturation magnetization of magnetite of $M_s=470~\mathrm{kA/m}$) instead of the measured value of about 18. For the 25~nm core diameter  the measured $\chi^\prime$  value is 43  and the theoretical $\chi^\prime$ value is about 544. 
These large discrepancies are consistent with results reported in the literature on the same kind of samples \cite{andersson_13}. There are at least two possible reasons for this unexpected result. A fraction of the nanoparticles might be aggregate with the consequence of a reduced response. Secondly, the theoretical estimates are based on the bulk value susceptibility, while important deviations may occur in the synthesis procedure at the surface region of the nanoparticles, leading to an important decrease of the effective core diameter of the nanoparticles.


The sensitivity of the susceptometer in terms of standard deviation for a given $\mathrm{Fe_3O_4}$ mass was evaluated.  The standard deviation is a complex function of the operation frequency. An increase in its value (by a factor of two) is present at low frequencies (below 100~Hz), which is attributed to the 1/f noise, and at frequencies above 10~kHz. In the intermediate frequencies a standard deviation of $4\times10^{-3}~\mathrm{mg}/M_{sample}$ for $\chi^\prime$ and $1\times10^{-2}~\mathrm{mg}/M_{sample}$ for $\chi^{\prime\prime}$  is found, where $M_{sample}$ is the mass of the $\mathrm{Fe_3O_4}$ sample. With respect to the instrumental limit evaluated in Sec.\ref{sec:setup}, these values are 6 and 15 times larger, respectively.

\section{Conclusion}
A  susceptometer is built using only standard and basic laboratory instrumentation. A commercial DAQ card is used to acquire triggered data samples from which both the magnetizing field and the magnetization signals are inferred. In contrast to the most common approaches, based on phase-sensitive detection with lock-in amplifiers, here the phase is evaluated with the help of a single-tone-extraction algorithm. The susceptometer works at room temperature and at low magnetic field values (of the order of a few 1~mT around 10~Hz, and down to $20\mathrm{\mu T}$ at 100~kHz). Its frequency response is calibrated using samples made of Cu wires, whose theoretical behavior is recalled and discussed. The susceptometer is used to determine the $\chi^\prime$ and $\chi^{\prime\prime}$ of samples containing mono-dispersed MNPs. These measurements make it possible to determine the frequency range in which the superparamagnetic behaviour of magnetic nanoparticles occurs.

\acknowledgments
This work was supported to a large extent by the national project FIRB RBAP11ZJFA -005 financed by the Italian Ministry for Education, University and Research. One author (A.~K.) is pleased to acknowledge the partial support provided through the project MagNanoP/POR-CRO-FSE-2007-2013-Asse-IV financed by the Region of Tuscany. The authors are indebted to Profs.~R.~Barbucci, G.~Marletta, A.~M.~Atrei and all the colleagues on the FIRB project for the useful discussions and the fruitful collaboration. They thank L.~Stiaccini and C.~Stanghini  for their valuable technical support, and E.~Thorley of Language Box (Siena) for helping to improve the manuscript.

\appendix
\section{The susceptibility of copper wires}

The result quoted in the main text (see Eq.\eqref{eq:chi_rame}) makes use of Bessel functions of a complex argument. These functions are not widely available  from a
numerical point of view. However the expression for  susceptibility can be translated into a  simpler and more accessible form using the Kelvin
functions \cite{abramowitz:stegun}. In fact they are defined as:
\begin{eqnarray}
  \label{eq:def:ber:bei}
  \ber_{\nu}(x) + i\, \bei_{\nu}(x) \equiv
  J_{\nu}\left( \cfrac{x}{\sqrt{2}}( -1 + i) \right) = \\ \nonumber
  \e^{i\,\pi\nu} J_{\nu}\left( \cfrac{x}{\sqrt{2}}(1-i) \right)
\end{eqnarray}
where $\nu  \geq 0$,  $x \geq  0$ and $J_{\nu}$  are the  usual Bessel functions of the first kind.  Writing $z$ of \eqref{eq:chi_rame} as  $z = x\,(1-i)/\sqrt{2} $ the 
theoretical susceptibility becomes:
\begin{equation}
  \label{eq:the:susc}
  1 + \chi_{th} =  - \cfrac{\sqrt{2}\,(1+i)}{x}\; \cfrac{\ber_1(x) + i
    \bei_1(x)}{\ber_0(x) + i \bei_0(x)}
\end{equation}
which, after some algebra using  the  formula 9.9.16  and 9.9.18  of  \cite{abramowitz:stegun}, can be cast as: 
\begin{equation}
  \label{eq:chi:fin}
  \chi_{th} = \left( -1 + \frac{2}{x} \frac{q_0}{p_0} \right) 
- i \, \frac{2}{x} \frac{r_0}{p_0}
\end{equation}
where  the $p_0$,  $q_0$ and  $r_0$  are particular  instances of  the cross-products: 
\begin{eqnarray*}
  p_{\nu}(x) &=& \ber_{\nu}^2(x) + \bei_{\nu}^2(x)\\
  q_{\nu}(x)      &=&       \ber_{\nu}(x)\,      \bei_{\nu}'(x)      -
  \ber_{\nu}'(x)\bei_{\nu}(x)\\
  r_{\nu}(x) &= &       \ber_{\nu}(x)\,      \ber_{\nu}'(x)      +
  \bei_{\nu}'(x)\bei_{\nu}(x).
\end{eqnarray*}

The functions  $p_0$, $q_0$  and $r_0$ can  be defined from their power series:
\begin{eqnarray}
\label{eq:p0q0r0_smallx}
  p_0 &= & \sum_{k=0}^{\infty} \cfrac{(x^2/4)^{2k}}{(2k)!\,k!^2}  \nonumber \\
q_0 &=& \frac{x}{2} \sum_{k=0}^{\infty} \cfrac{(x^2/4)^{2k}}{(2k+1)!\,k!^2} \\
r_0  &=&  \frac{x^3}{8}  \sum_{k=0}^{\infty}\cfrac{(x^2/4)^{2k}}{(2k+2)!\,(k+1)!\,k!}  \nonumber
\end{eqnarray}
which converge for any $x$, but are not suitable for numerical work if $x$  is large.  However  the asymptotic  behaviours for  $x\rightarrow
\infty$ are also known \cite{abramowitz:stegun}:
\begin{eqnarray}
	p_0 &=&  \cfrac{\e^{x\sqrt{2} }}{ 2\pi\,x }  \times \left( 1 + \cfrac{1}{4\rd} \cfrac{1}{x}  + \cfrac{1}{64} \cfrac{1}{x^2}+ \right . \nonumber \\
	&   & \left . - \cfrac{33}{256\rd} \cfrac{1}{x^3} - \cfrac{1794}{8192} \cfrac{1}{x^4} +O(x^{-5}) \right)  \nonumber \\
  q_0 &=& \cfrac{\e^{x\sqrt{2} }}{ 2\pi\,x}  \times \left( \cfrac{1}{\rd} + \cfrac{1}{8} \cfrac{1}{x}  + \cfrac{9}{64\rd} \cfrac{1}{x^2}+   \right . \label{eq:p0q0r0_largex}\\
	&   &   \left. + \cfrac{39}{512} \cfrac{1}{x^3} + \cfrac{75}{8192\rd} \cfrac{1}{x^4} +O(x^{-5}) \right) \nonumber \\
  r_0 &=& \cfrac{\e^{x\sqrt{2} }}{ 2\pi\,x}   \times  \left( \cfrac{1}{\rd} - \cfrac{3}{8\rd} \cfrac{1}{x}  - \cfrac{15}{64\rd} \cfrac{1}{x^2} + \right . \nonumber \\
	&  &  \left . - \cfrac{45}{512} \cfrac{1}{x^3} + \cfrac{315}{8192\rd} \cfrac{1}{x^4} +O(x^{-5}) \right). \nonumber 
\end{eqnarray}

Using these  expressions the leading  terms for the susceptibility ($\chi = \chi^\prime- i\,\chi^{\prime\prime}$)  can be derived for small $x$:
\begin{eqnarray*}
  \chi' & = & -{\frac {1}{48}}{x}^{4}+{\frac {19}{30720}}{x}^{8}-{\frac {229}{
      12386304}}{x}^{12}+O \left( {x}^{16} \right)  \\
\chi'' & = & {\frac {1}{8}}{x}^{2}-{\frac {11}{3072}}{x}^{6}+{\frac {473}{4423680}
}{x}^{10}+O \left( {x}^{14} \right)
\end{eqnarray*}
and large $x$:
\begin{eqnarray*}
  \chi' &=& -1+{\frac {\sqrt {2}}{x}}+{\frac { \sqrt {2}}{8 {x}^{3}}}+\frac{1}{4\,{x}^{
4}}+{\frac {25}{128}}\,{\frac {\sqrt {2}}{{x}^{5}}}+O \left( {x}^{-6}
 \right) \\
\chi'' &=& {\frac {\sqrt {2}}{x}}-\frac{1}{{x}^{2}}-{\frac {\sqrt {2}}{8{x}^{3}}}+{
\frac {25}{128}}\,{\frac {\sqrt {2}}{{x}^{5}}}+O \left( {x}^{-6}
 \right). 
\end{eqnarray*}
These series expansions of $\chi$ may help in visualizing the asymptotic behaviour, but are not favourable for a numerical evaluation, as they would require too many terms to obtain adequate precision. A more favourable choice is based on the use of the rational expression Eq.(\ref{eq:chi:fin}), with appropriate evaluations of $p_0$, $q_0$ and $r_0$. 
The numerical evaluation of $p_0$, $q_0$ and $r_0$ can be performed  using the power series expressions  Eq.(\ref{eq:p0q0r0_smallx}) with 10  terms  for $x  \leq 10 $ and  the asymptotic expansions given in Eq.(\ref{eq:p0q0r0_largex}) for larger $x$. Of course computer algebra software such as Maple can also be used, but the   approximations reported above can be  more profitable  when the evaluation of those quantities has to be integrated into other kinds of  software, such as  programmes devoted to  data acquisition and instrumentation management.

\begin{figure}[htbp] \centering
\vspace{12pt}
\includegraphics [width=8cm] {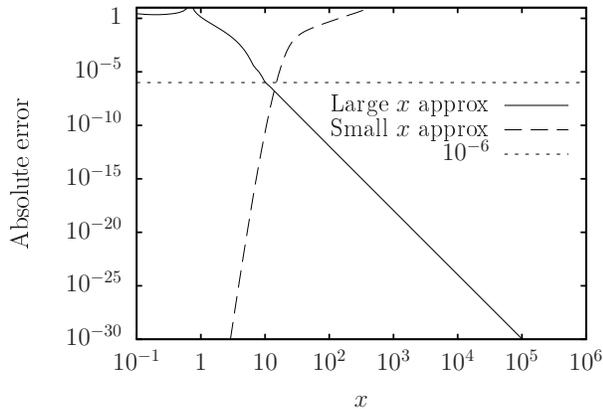}
\caption{Comparison of absolute  errors introduced by the use of two numerical, approximated evaluations of the susceptibility. The difference between the exact value and two kinds of estimates is reported. The dashed line describes the error obtained by a truncation of the sum Eq.(\ref{eq:p0q0r0_smallx}) at the 10$^{th} $term, while the solid line describes the error introduced by the five-term asymptotic approximation described in Eq.(\ref{eq:p0q0r0_largex})}
\label{fig:err}
\end{figure}

In Fig.~\ref{fig:err}  the absolute error  $| \chi_{exact} - \chi_{approx}  |$ is plotted as a function of $x$, where $\chi_{exact}$ is calculated
using Maple, while  $\chi_{approx}$ is calculated using the small $x$  (Eq.(\ref{eq:p0q0r0_smallx})) and large $x$ (Eq.(\ref{eq:p0q0r0_largex})) approximations, respectively. As shown, the method leads to a maximum absolute error as small as $10^{-6}$.

\





\begin{thebibliography}{10}

\bibitem{chen_04}
D.-X. {Chen}, ``High-field ac susceptometer using {H}elmholtz coils as a
  magnetizer,'' {\em Measurement Science and Technology}, vol.~15, no.~6,
  pp.~1195--1202, 2004.

\bibitem{hergt_06}
{R. Hergt}, {S. Dutz}, {R. M\"uller}, and {M. Zeisberger}, ``Magnetic particle
  hyperthermia: Nanoparticle magnetism and materials development for cancer
  therapy,'' {\em Journal of Physics Condensed Matter}, vol.~18, no.~38,
  pp.~S2919--S2934, 2006.

\bibitem{mamiya_11}
H.~Mamiya and B.~Jeyadevan, ``Hyperthermic effects of dissipative structures of
  magnetic nanoparticles in large alternating magnetic fields,'' {\em
  Scientific Reports}, vol.~1, 2011.

\bibitem{arruebo_07}
{Manuel Arruebo}, {Rodrigo Fern\'andez-Pacheco}, {M. Ricardo Ibarra}, and
  {Jes\'us Santamar\'ia}, ``Magnetic nanoparticles for drug delivery,'' {\em
  Nano Today}, vol.~2, no.~3, pp.~22 -- 32, 2007.

\bibitem{grossman_04}
H.~L. Grossman, W.~R. Myers, V.~J. Vreeland, R.~Bruehl, M.~D. Alper, C.~R.
  Bertozzi, and J.~Clarke, ``Detection of bacteria in suspension by using a
  superconducting quantum interference device,'' {\em Proceedings of the
  National Academy of Sciences of the United States of America}, vol.~101,
  no.~1, pp.~129--134, 2004.

\bibitem{hergt_98}
R.~Hergt, W.~Andra, C.~d'Ambly, I.~Hilger, W.~Kaiser, U.~Richter, and H.-G.
  Schmidt, ``Physical limits of hyperthermia using magnetite fine particles,''
  {\em IEEE Transactions on Magnetics}, 1998.

\bibitem{couach_85}
M.~Couach, A.~Khoder, and F.~Monnier, ``Study of superconductors by a.c.
  susceptibility,'' {\em Cryogenics}, vol.~25, no.~12, pp.~695 -- 699, 1985.

\bibitem{laurent_08}
P.~{Laurent}, J.~F. {Fagnard}, B.~{Vanderheyden}, N.~H. {Babu}, D.~A.
  {Cardwell}, M.~{Ausloos}, and P.~{Vanderbemden}, ``{An ac susceptometer for
  the characterization of large, bulk superconducting samples},'' {\em
  Measurement Science and Technology}, vol.~19, p.~085705, Aug. 2008.

\bibitem{dumelow_01}
T.~{Dumelow}, M.~M. {Xavier}, F.~A.~O. {Cabral}, and C.~{Chesman}, ``{A simple
  AC susceptometer mounted on a cryostat cold finger},'' {\em Journal of
  Magnetism and Magnetic Materials}, vol.~226, pp.~2063--2064, May 2001.

\bibitem{chen_99}
D.-X. Chen, L.~Pascual, E.~Fraga, M.~Vazquez, and A.~Hernando, ``Magnetic and
  transport eddy-current anomalies in cylinders with core-and-shell regions,''
  {\em Journal of Magnetism and Magnetic Materials}, vol.~202, no.~23, pp.~385
  -- 396, 1999.

\bibitem{chen_05}
D.-X. Chen and C.~Gu, ``Ac susceptibilities of conducting cylinders and their
  application in electromagnetic measurements,'' {\em IEEE Transactions of
  Magnetics}, vol.~41, no.~9, pp.~2436--2446, 2005.

\bibitem{goldfarb_84}
R.~Goldfarb and J.~Minervini, ``Calibration of ac susceptometer for cylindrical
  specimens,'' {\em Review of Scientific Instruments}, vol.~55, pp.~751--764,
  1984.

\bibitem{grandke_83}
T.~Grandke, ``Interpolation algorithms for discrete fourier transforms of
  weighted signals,'' {\em IEEE Trans. Instrum. Meas.}, vol.~32, pp.~350--355,
  1983.

\bibitem{lanczos_56}
C.~Lanczos, {\em {Applied Analysis, paragraphs III.5; IV.22}}.
\newblock Englewood Cliffs, N.J. Prentice-Hall, 1956.

\bibitem{chen_10}
D.-X. {Chen} and V.~{Skumryev}, ``Calibration of low-temperature ac
  susceptometers with a copper cylinder standard,'' {\em Review of Scientific
  Instruments}, vol.~81, no.~2, p.~025104, 2010.

\bibitem{landau_84}
L.~D. Landau, L.~P. Pitaevskii, and E.~M. Lifshits, {\em Electrodynamics of
  continuous media 2nd ed.}
\newblock Pergamon, Oxford, 1984.

\bibitem{andersson_13}
M.~Andersson, {\em Modeling and characterization of magnetic nanoparticles
  intended for cancer treatment}.
\newblock PhD thesis, Uppsala University, Solid State Physics, 2013.

\bibitem{kotitz_95}
{R. K\"otitz}, {P.C. Fannin}, and {L. Trahms}, ``Time domain study of
  {B}rownian and {N}{\'e}el relaxation in ferrofluids,'' {\em Journal of
  Magnetism and Magnetic Materials}, vol.~149, no.~12, pp.~42 -- 46, 1995.
\newblock Proceedings of the Seventh International Conference on Magnetic
  Fluids.

\bibitem{abramowitz:stegun}
M.~Abramowitz and I.~A. Stegun, {\em Handbook of Mathematical Functions with
  Formulas, Graphs, and Mathematical Tables}.
\newblock New York: Dover, 1964.

\end{thebibliography}

\end{document}